\documentclass[floatfix,twocolumn,showpacs,preprintnumbers,amsmath,amssymb,superscriptaddress]{revtex4}
\usepackage{graphicx,psfrag}
\usepackage{amsfonts}

\def\simlt{\lower.5ex\hbox{$\; \buildrel < \over \sim \;$}}
\def\simgt{\lower.5ex\hbox{$\; \buildrel > \over \sim \;$}}
\def\simpropto{\lower.2ex\hbox{$\; \buildrel \propto \over \sim \;$}}

\newcommand{\newc}{\newcommand}
\newc{\gsim}{\lower.7ex\hbox{$\;\stackrel{\textstyle>}{\sim}\;$}}
\newc{\lsim}{\lower.7ex\hbox{$\;\stackrel{\textstyle<}{\sim}\;$}}
\newc{\gev}{\,{\rm GeV}}
\newc{\mev}{\,{\rm MeV}}
\newc{\ev}{\,{\rm eV}}
\newc{\kev}{\,{\rm keV}}
\newc{\tev}{\,{\rm TeV}}

\newc{\mz}{M_Z}
\newc{\mpl}{M_*}
\newc{\mw}{m_{\rm weak}}
\newc{\nr}[1]{N^c_R{}_{#1}}
\usepackage{amsmath}
\def\beq{\begin{equation}}
\def\eeq{\end{equation}}
\def\bea{\begin{eqnarray}}
\def\eea{\end{eqnarray}}
\def\bitem{\begin{itemize}}
\def\eitem{\end{itemize}}
\newc{\ie}{{\it i.e.}}          \newc{\etal}{{\it et al.}}
\newc{\eg}{{\it e.g.}}          \newc{\etc}{{\it etc.}}
\newc{\cf}{{\it c.f.}}
\def\abs#1{\left| #1\right|}

\def\inv{^{\raise.15ex\hbox{${\scriptscriptstyle -}$}\kern-.05em 1}}
\def\lbar{{\lower.35ex\hbox{$\mathchar'26$}\mkern-10mu\lambda}} 

\let\<=\langle
\let\>=\rangle

\let\+=\uparrow

\let\al=\alpha
\let\be=\beta
\let\ga=\gamma

\let\De=\Delta

\let\La=\Lambda

\let\si=\sigma

\let\th=\theta

\begin{document}
\thispagestyle{empty}

\begin{center}
\title{Emergent Flux from Particle Collisions Near a Kerr Black Hole}
\date{October 13th 2010}
\author{M\'aximo Ba\~nados}
\email[]{maxbanados@fis.puc.cl}
\affiliation{Facultad de F\'{\i}sica, Pontificia Universidad Cat\'{o}lica de Chile, Av. Vicu\~{n}a Mackenna 4860, Santiago, Chile }
\author{Babiker Hassanain}
\email[]{b.hassanain1@physics.ox.ac.uk}
\affiliation{Rudolf Peierls Centre for Theoretical Physics, University of Oxford, 1 Keble Rd., Oxford OX1 3NP, UK}
\author{Joseph Silk}
\email[]{j.silk1@physics.ox.ac.uk}
\affiliation{Physics Department, University of Oxford, Oxford, OX1 3RH, UK}
\author{Stephen M. West}
\email[]{stephen.west@rhul.ac.uk}
\affiliation{Royal Holloway, University of London, Egham, TW20 0EX, UK}
\affiliation{Rutherford Appleton Laboratory, Chilton, Didcot, OX11 0QX, UK}

\begin{abstract}
The escape fraction at infinity is evaluated for massless particles produced in collisions of weakly interacting particles accreted into a density spike near the particle horizon of an extremal Kerr black hole, for the case of equatorial orbits. We compare with the Schwarzschild case, and argue that in the case of extremal black holes, redshifted signatures can be produced that could potentially explore the physics of particle collisions at centre of mass energies that extend beyond those of any feasible terrestrial accelerator.
 \end{abstract}
\maketitle
 
\end{center}

\section{Introduction}
\label{intro}

Some of us have recently argued that rotating black holes  surrounded by relic cold Dark Matter (DM) density spikes may act as particle accelerators \cite{banados}.  In the limit of maximally rotating, extremal Kerr black holes, we showed that collisions between particles, e.g. weakly interacting massive DM particles, may reach arbitrarily high center-of-mass  energies. For some related works see \cite{Piran:1977dm,Baushev,jacobson, berti,grib,Wei:2010vc,Lake:2010bq}. 

In this sequel, we compute the escape fraction and flux at infinity of the highly blue-shifted particles. 
We  demonstrate that the escape fraction is finite and of interest for
any value of $a,$ although the price one pays for sub-extremality is that the
achievable Centre-of-Mass Frame (CMF) collision energy is limited.

Our earlier work attracted considerable comment with regard to the fact that extremality for a Kerr black hole might not be achievable in nature
\cite{jacobson, berti}, although not all authors concur on this point \cite{grib}.  Accretion spins up the hole but radiative back-reaction limits the degree of spin-up. The limit
$a\simlt 0.9980\pm 0.0002$ was derived \cite{Thorne} for a thin accretion disk, and criticised as  being nonrealistic for  more general situations \cite{abramowicz1980}. More recently, the role of binary black hole mergers has been studied
and superradiant extraction  of angular momentum from the larger black hole has been shown to set the limit $a\simlt 0.9979\pm 0.0001$ \cite{kesden2010}. However we note that these limits apply only to astrophysical
constraints, and that string theory may provide alternative options for generating extremal black holes.

Given these motivations, it is imperative that we obtain an estimate for the flux expected from the ultra-energetic collisions that take place in the near-horizon region of Kerr black holes \cite{banados}. Concretely, we will 
focus on the flux emitted by DM spikes that form around Intermediate-Mass Black Holes (IMBHs), as detailed in \cite{BZS2005}. Such DM spikes are expected to extend down to distances that are of order 
the Schwarzschild horizon radius, and to have a density that is essentially set by the annihilation rate of the DM. We take into account the fact that massless products of ultra-energetic collisions may be captured by the black hole by including a fully-relativistic escape function into the computation. 

The paper is organised as follows: we first set the scene in the next section, and present a relativistic formula for the total flux expected at infinity from collisions around an IMBH. This formula requires knowledge of the escape function, which encodes the proportion of massless products that arrive at infinity, as well as the DM density profile around the black hole. We discuss the escape function in Section~\ref{escapefraction}, and the density profile in Section~\ref{density}. We present our results in Section~\ref{results}, and end by expanding on some open issues that must be tackled to improve and extend the results of this paper.

\section{The annihilation rate, density profile and escape fraction.}
\label{annihilation}

Our aim is to obtain the flux emanating from particle annihilations in a DM spike collapsing around a rotating black hole.
We first explain how we intend to compute the flux, for steady-state (time-independent) conditions. The first ingredient is the DM
number density in the vicinity of the black hole. We model this using a distribution function in phase space $n(\vec{x},\vec{v})$,
where $\vec{x}$ and $\vec{v}$ are the coordinates of real space and velocity space respectively.  This distribution function $n(\vec{x},\vec{v})$, which
is a Lorentz scalar, enters the annihilation rate quadratically. Defining $\sigma(\vec{x},\vec{v}_1,\vec{v}_2)$ as the annihilation cross section for a pair of DM particles and $v_{rel}(\vec{x},\vec{v}_1,\vec{v}_2)=\sqrt{1-\left(\frac{1}{g_{\mu\nu}U_1^\mu U_2^\nu}\right)^2}$ as the
relative velocity of two DM particles with velocities $U_1$ and $U_2$, we propose the following formula for the {\emph{invariant cross section}}
\begin{widetext}\
\begin{equation}\label{annihilation}
dA= \sigma(\vec{x},\vec{v}_1,\vec{v}_2) \, v_{rel}(\vec{x},\vec{v}_1,\vec{v}_2) \, g_{\mu\nu}U_1^\mu U_2^\nu \, . n(\vec{x}_1,\vec{v}_1). \,  n(\vec{x}_2,\vec{v}_2) d{\mathcal{V}}_{1}d{\mathcal{V}}_{2} \, ,
\end{equation}
\end{widetext}
where $d{\mathcal{V}}_{1,2}$ denotes the velocity space of the colliding particles. The preceding equation tells us the number of annihilations taking place locally. Our objective is of course to find the flux of massless particles escaping the gravitational field of the black hole. Only massless particles produced with momenta satisfying certain conditions will propagate out to infinity. To obtain the number of escaping massless particles from the annihilation rate, we must therefore dress the rate of Eq.~(\ref{annihilation}) with an ``escape function'' which
encodes the probability that the massless annihilation products escape to infinity. We expect this function to go to zero at the horizon (such that all of the annihilation products
get captured there) and approach one at infinity, so that all of the annihilation products very far from the black hole escape to infinity and contribute to the flux. Call the escape function
$e(\vec{x},\vec{v}_1,\vec{v}_2)$, where we make explicit its dependence on the velocities of the colliding DM particles. Folding this in with the annihilation rate above, we
have that the number of massless particles escaping the black hole per unit time $N_{\infty}$ is given by
\begin{equation}\label{flux}
N_{\infty}=\int_{{\mathcal{V}}_1}\int_{{\mathcal{V}}_2}\int_{V} e(\vec{x},\vec{v}_1,\vec{v}_2) \,  dA \, \sqrt{-g} \,d^3\, x \, .
\end{equation}
Note that in principle we can use a similar formulation to obtain the flux spectrum, but our sole purpose is to show that the flux is potentially observable.
If the observer is a distance $D$ away from the black hole, then to find the flux received per unit area per unit time by the observer we must divide the above by $4 \pi D^2$ as usual. Note 
that in this work we limit the phase space integration domains ${\mathcal{V}}_{1,2}$ to in-falling DM particles moving on geodesics that are captured by the black hole, the so-called ``plunge" orbits,
as those are the ones which lead to collisions with high CMF energies. For our purposes, the spatial integral will be computed in a spherical volume $V$ around the black hole,
extending to a radius $r_f$, measured in units where the Schwarzschild radius $R_s=m_{BH}G/c^2$ is set to 1. The upper limit of the volume integral is a free variable (in principle), and we may choose to restrict it in
order to focus on the size of the flux emerging from the near horizon region, where the CMF energies of the collisions are high.

A crucial ingredient required to calculate the flux is the density profile of DM around the black hole. In this paper we use a density profile corresponding to the ``spike" proposed in \cite{Gondolo:1999ef, BZS2005}. This density profile includes the effects of DM annihilation on the DM density (as well as other effects) and we shall make the assumption that it provides a good description in all regions down to the black hole horizon. We stress that this density profile is also non-relativistic, and so its convolution with a fully relativistic (and exact) escape function is a source of potential inconsistency but this method provides a relatively straight forward way to estimate the density of DM particles around a black hole. Finally, the types of density spikes described in \cite{Gondolo:1999ef} correspond to density profiles around non-rotating black holes. The more energetic collisions occur for rotating black holes, and therefore in a more detailed analysis aimed at finding exact numbers a relativistic density profile is needed. Again we stress that we are only interested at this stage in an order of magnitude estimate for the size of the emergent flux so we leave such refinements to future investigations. 

Regardless of how one estimates the density profile, in order to calculate the emergent flux we need to know what fraction of collision products escape the gravitational field of the black hole. To make the estimate as straightforward and as clear as possible, we only consider the simplified case of the annihilation of two DM particles into two massless particles (e.g. two neutrinos or two photons). The escaping fraction of collision products is an $r$-dependent quantity. For the case of a Schwarzschild black hole the escape fraction as a function of $r$ is a known result and a useful treatment can be found in \cite{mtw}. 

When considering the case of plunge orbits, the escape fraction becomes more complicated due to the fact that the collision products are boosted towards the black hole. In the next section, we describe the calculation of the escape fraction as a function of $r$ for both Schwarzschild and Kerr black holes, including the relativistic effects of the boosts.

\section{Escape Function}
\label{escapefraction}

In this section, we present an analytic treatment of the escape fraction for massless particles. These particles are produced in the collisions of the DM particles and are consequently highly boosted towards the centre of the black hole. The details of this boost are dependent on the momentum of the DM particles. This of course is determined by the angular momentum of the two DM particles. It is a relatively straightforward exercise to construct the escape function for a massless particle produced in the gravitational field of a Schwarzschild black hole \cite{mtw}. Here we want to do a similar calculation but with the massless particles boosted according to the kinematics of the initial colliding DM particles.

Let us first examine the motion of massless particles in the Locally-NonRotating-Frame (LNRF), which is the frame in which an observer who rotates with the local geometry does their measurements. 
Working in the equatorial plane, the LNRF momentum of a massless particle is given by 
\beq
P^{LNRF}=\left(P^{LNRF}_{\hat{t}}, P^{LNRF}_{\hat{r}}, P^{LNRF}_{\hat{\th}}\right)
\eeq
where the components take the forms  \cite{BPT}
\begin{widetext}
\bea
P^{LHRF}_{\hat{t}}(r, a)&=&\sqrt{1+\frac{2}{r}+\frac{4}{a^2+(r-2) r}},\nonumber\\
P^{LHRF}_{\hat{r}}(r, a, \si, b)&=&\sigma\;\frac{\sqrt{r^3+a^2 (2+r)} \sqrt{1+\frac{2}{r}+\frac{4}{a^2+(r-2) r}} \sqrt{r^3+2 (a-b )^2+r (a^2-b^2)}}{r^3+a^2 (2+r)-2 a b},\nonumber\\
P^{LHRF}_{\hat{\th}}(r, a, b)&=&\frac{r b\sqrt{a^2+(r-2) r} \sqrt{1+\frac{2}{r}+\frac{4}{a^2+(r-2) r}}  }{r^3+a^2 (2+r)-2 a b},
\eea
\end{widetext}
where we are working in units where the mass of the black hole and the Schwarzschild radius $R_s$ are set equal to 1.
In this frame we can establish the conditions on the massless particle's 3-momentum as a function of $r$ such that the it escapes the gravitational field of the black hole. This is most easily performed by analysing the turning points (i.e. where $\dot{r}=0$) of the massless particle as a function of $r$ and $b$.

The equation of motion for $\dot{r}$ on the equatorial plane is
\beq
\dot{r}=1-\frac{(b^2-a^2)}{r^2}+\frac{2(b-a)^2}{r^3}.
\label{rdot}
\eeq

Setting this equal to zero and solving for $b$ as a function of $r$ we find the two solutions
\beq
b(r)=\frac{-2a\pm\sqrt{a^2r^2-2r^3+r^4}}{r-2}.
\label{solb}
\eeq

We can now use these solutions to tell us whether a massless particle produced at some distance $r$, with particular values of $b$ and $\si$ ($\si=\pm1$ and determines whether the massless particle is moving to larger $(+1)$ or smaller $(-1)$ radii) will escape the black hole.

To do this for general $a$ analytically is tricky, and we choose to investigate two values of $a$, namely $a=0$, the Schwarzschild limit, and the extreme Kerr case of $a=1$. Other values of $a$ give results that are numerically not too different from the analytic results presented below and we leave the details of a numerical treatment for arbitrary $a$ to future work \cite{williams}. Our goal here is to find an estimate for the emergent flux rather than to give a precise prediction.

\begin{figure}[h!]
\includegraphics[scale=0.75]{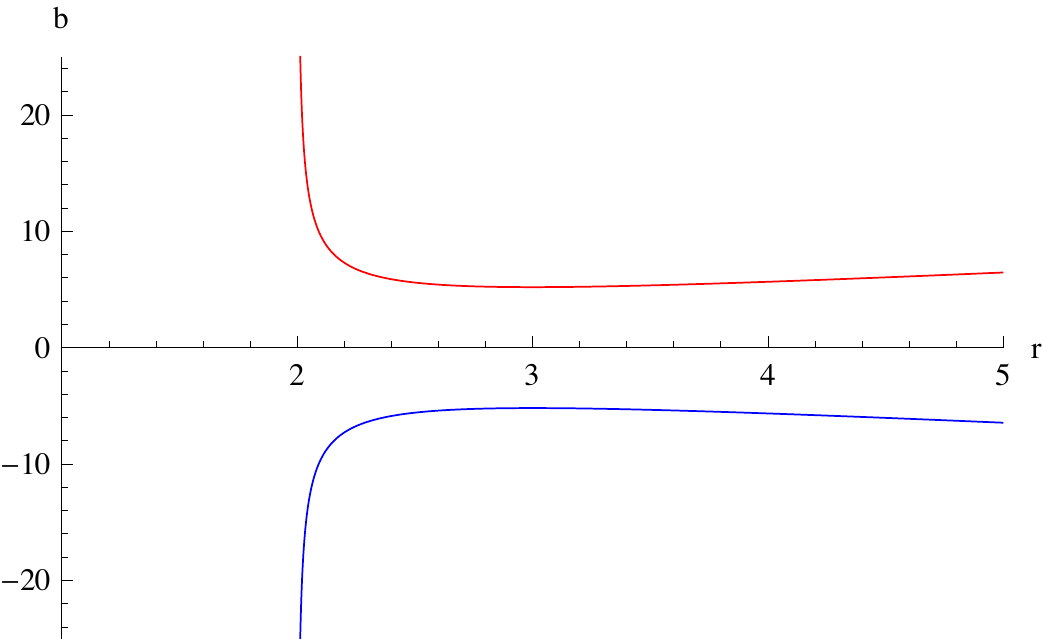}
\caption{A plot of turning points for Schwarzschild black holes in terms of $b$ as a function of $r$ (in units where $R_s=1$). The upper (red) curve is $b_+^{a=0}(r)$ whereas the lower (blue) curve is $b_-^{a=0}(r)$.\label{bvsr}}
\end{figure}

Starting with the Schwarzschild case, $a=0,$ the two solutions of Eq.~\ref{solb} take the forms
 \beq
b^{a=0}_{\pm}(r)= \pm\frac{r^{3/2}}{\sqrt{r-2}} \, .
 \eeq
We will for the remainder of the Schwarzschild discussion omit the superscript $a=0$, to avoid clutter.
By plotting these solutions in FIG.~\ref{bvsr} we see that for the positive solution there is a minimum and for the negative solution there is a maximum. Both these extrema occur at $r=3$ and have values $b_+(3)=3\sqrt{3}$ and $b_-(3)=-3\sqrt{3}$. We can now read off the conditions on $b$ as a function of $r$ that determine whether the massless particle will escape or be captured by the black hole.

The conditions for the massless particle to escape are for
\bea
\noindent
r<3: \si=+1\; {\rm and}\;-3\sqrt{3} >b >3\sqrt{3} \, ,
\label{con1}
\eea
which simply reflects the fact that a particle produced in the region $r<3$, with a value of $b$ in between the minimum and maximum, will move towards infinity unhindered by the potential barrier. If the particle were 
produced with $b> 3\sqrt{3}$, then it will hit the potential barrier and bounce back into the black hole. 
Similar reasoning then yields the following additional conditions for escape, which apply in the region $r>3$:
\bea\label{con2}
 r>3 : \si&=&+1\; {\rm and}\; b_-(r) >b > b_+(r);\\\label{con3}
\si &=&-1\; {\rm and}\; 3\sqrt{3}>b >b_+(r); \\\label{con4}
\si &=&-1\;{\rm and}\; b_-(r)>b > -3\sqrt{3}.
\eea

For the $\si=-1$ conditions, the massless particle will initially be moving to smaller $r$ but will reach a turning point (at a distance $r$ determined by the value of $b(r)$) and will be reflected by the effective potential leading to the escape of the massless particle.

To calculate the escape function, we apply the following logic. We assume that the annihilation of two DM particles proceeds as $\chi\chi \rightarrow p_1p_2,$ where we assert that the $p_i$s are massless particles. As an estimate designed to provide a reasonable order of magnitude measure, we only concern ourselves with the question of whether $p_1$ escapes. By doing this, we avoid the complications introduced by trying to consistently analyse whether either, both or neither $p_1$ or $p_2$ escape.

We construct the measure by assuming that in the CMF of the collision, $p_1$ is produced isotropically, that is, there is no preferred direction. In the CMF, there is an $r$ dependent range of angles that correspond to particular values of $b$ and $\si$ (evaluated in the LNRF) that satisfy the conditions for $p_1$ to escape. In other words, the conditions on the values of $b$ and $\si$ for the massless particle to escape in the LNRF (Eqs.~(\ref{con1}) - (\ref{con4})) can be mapped onto a range of angles in the CMF of the collision in which $p_1$ is produced. Because of the assumption that the production in the CMF is isotropic, the escape fraction can be computed in terms of the ratio of this range of angles over the full $2\pi$ in the CMF on the equatorial plane. The situation can therefore be summarised as follows: two DM particles collide at a given $r$ with two respective angular momenta. We have two frames: the LNRF frame, which we can think of conventionally as the lab frame, and the CMF of the collision. The two frames are related by a boost, which is defined by the radial position and the angular momenta of the colliding DM particles. It is this boost that provides the mapping between the escape conditions on $b$ and $\si$ and the range of angles in the CMF. 

One immediate important point is that by assuming the massless particles are only produced and escaping in the equatorial plane, we are potentially overestimating the escape fraction. However, this is not entirely clear as some particles could still escape even if they have trajectories that take them off the equatorial plane. For simplicity we restrict our analysis to the equatorial plane. In a paper to appear, a numerical analysis is performed, \cite{williams}, where it is shown that this approximation is indeed a good one to the accuracy to which we wish to work in this analysis. 

To find the $r$-dependent range of angles in the CMF that corresponds to the full range of escape conditions for $b$ and $\si$ in the LNRF, we can construct dot products between the 2 spatial components of the massless particle momentum vectors evaluated in the CMF (i.e. the $r$ and $\th$ components) corresponding to the extreme values of $b$ for the appropriate values of $\si$ and solve for the angle between them. This angle is then the escape range in the CMF, and all massless particles produced inside it escape. 

First we need to find the expressions for the momenta in the CMF. To do this we simply apply a boost to the LNRF momenta for the massless particles such that
\beq
P^{CMF}=\La.P^{LNRF},
\eeq
is the CMF momenta and where $\La$ is a general Lorentz boost in the equatorial plane that can be written as
\begin{multline}\nonumber
\La=\\\nonumber
\left(
\begin{array}{ccc}
 \gamma  & - \beta \gamma \cos \chi & -\sin\chi  \beta  \gamma  \\
 -\cos \chi \beta  \gamma  & 1+\cos \chi^2 (-1+\gamma ) & \cos \chi \sin\chi  (-1+\gamma ) \\
 -\sin\chi  \beta  \gamma  & \cos \chi \sin\chi  (-1+\gamma ) & 1+\sin\chi^2 (-1+\gamma )
\end{array}
\right),
\end{multline}
where $\chi$, $\ga$ and $\be$ are the usual parameters of the Lorentz boost. The exact form of this boost is determined by the momenta of the DM particles that are colliding together. Parameterising the momenta of the massive colliding particles (where we have set the mass of the DM particle $\mu=1$) as
\bea
Q_i(r, \al_i, a)&=&(Q_{i{\hat{t}}}, Q_{i\hat{r}}, Q_{i\hat{\th}})
\eea
where the momentum vectors are parameterised by the angular momentum $\al_i$ and we use $Q$ instead of $P$ to make it clear that this momentum vector is for a massive particle as opposed to a massless particle. The explicit forms after taking $a=0$ are \cite{BPT}
\bea
Q_{i\hat{t}}(r, a=0)&=&\frac{1}{\sqrt{1-\frac{2}{r}}},\\
Q_{i\hat{r}}(r, \al_i, a=0)&=& -\frac{\sqrt{\left(2 r^2+2 \alpha_i^2-r \alpha_i ^2\right)}}{r^{3/2}\sqrt{1-\frac{2}{r}}},\\
Q_{i\hat{\th}}(r, \al_i, a=0)&=& \frac{\al_i}{r}.
\eea
The boost parameters in terms of these momentum components are then found to be
\bea\nonumber
\beta &=&\frac{\left((Q_{i\hat{r}}+Q_{j\hat{r}})^2+(Q_{i\hat{\th}}+Q_{j\hat{\th}})^2\right)^{1/2}}{(Q_{i\hat{t}}+Q_{j\hat{t}}) },\\\nonumber
\gamma &=&\frac{Q_{i\hat{t}}+Q_{j\hat{t}}}{\sqrt{(Q_{i\hat{t}}+Q_{j\hat{t}})^2-(Q_{i\hat{r}}+ Q_{j\hat{r}})^2-({Q_{i\hat{\th}}}+Q_{j\hat{\th}})^2}},\\\nonumber
\cos \chi &=&\frac{Q_{i\hat{r}}+Q_{j\hat{r}}}{\sqrt{(Q_{i\hat{r}}+Q_{j\hat{r}})^2+(Q_{i\hat{\th}}+Q_{j\hat{\th}})^2}},\\\nonumber
\sin \chi &=&\frac{Q_{i\hat{\th}}+Q_{j\hat{\th}}}{\sqrt{(Q_{i\hat{r}}+Q_{j\hat{r}})^2+(Q_{i\hat{\th}}+Q_{j\hat{\th}})^2}}.\nonumber
\eea

We are now in a position to construct the dot products. Using the boosted momenta of the massless final state particles we can construct the dot products. Remember that this is in the CMF now. Explicitly we construct the vectors as
\beq\nonumber
\tilde{P}(r, b, \si)=\left[ P^{CM}_{\hat{r}}(r, b, \si, a=0),\; P^{CM}_{\hat{\th}}(r, b, \si, a=0)\right].
\eeq

With reference to the conditions set out in Eq.~(\ref{con1}) for $r<3$ the range of escape angles, $\De \phi$, in the CMF reads
\beq\nonumber
\De\phi^{r<3}(r)=\abs{\arccos\;\left[\frac{\tilde{P}(r, \;3\sqrt{3}, \;+1).\tilde{P}(r, \;-3\sqrt{3}, \;+1)}{\abs{\tilde{P}(r, \;3\sqrt{3}, \;+1)}\abs{\tilde{P}(r, \;-3\sqrt{3}, \;+1)}}\right]},
\eeq
with the escape fraction for $r<3$, $EF^{r<3}$, given by
\beq
EF^{r<3}(r)=\frac{\De\phi(r)^{r<3}}{2\pi}.
\eeq
We again state that we are only considering motion in the equatorial plane. The corresponding expressions for $r>3$ are (following the conditions stated in Eqs.~(\ref{con2}, \ref{con3}, \ref{con4}))
\begin{multline}\nonumber
\De\phi^{r>3} (r)=\\
\abs{\arccos\;\left[\frac{\tilde{P}(r, \; b_+(r), \;+1).\tilde{P}(r, \;b_-(r), \;+1)}{\abs{\tilde{P}(r, \;b_+(r), \;+1)}\abs{\tilde{P}(r, \;b_-(r), \;+1)}}\right]}\\
+ \abs{\arccos\;\left[\frac{\tilde{P}(r, \;b_+(r), \;-1).\tilde{P}(r, \;3\sqrt{3}, \;-1)}{\abs{\tilde{P}(r, \;b_+(r), \;-1)}\abs{\tilde{P}(r, \;3\sqrt{3}, \;-1)}}\right]}
\\
+\abs{\arccos\;\left[\frac{\tilde{P}(r, \;-3\sqrt{3}, \;-1).\tilde{P}(r, \;b_-(r), \;-1)}{\abs{\tilde{P}(r, \;-3\sqrt{3}, \;-1)}\abs{\tilde{P}(r, \;b_-(r), \;-1)}}\right]}
\end{multline}
with the escape fraction for $r>3$, $EF^{r>3}$, given by
\beq
EF^{r>3}(r)=\frac{\De\phi(r)^{r>3}}{2\pi}.
\eeq
Putting this all together, the total escape fraction is 
\bea\nonumber
EF_{a=0}(r)&=&\frac{1}{2\pi}\left[\De\phi_{a=0}^{r<3}(r)\Theta(r-3)\right.\\
&&\left.+\De\phi_{a=0}^{r>3}(r)(1-\Theta(r-3))\right],
\eea
where $\Theta(r-3)=1$ for $r<3$ and zero otherwise. 

\begin{figure}[h!]
\vspace{0cm}
\centerline{\includegraphics[width=8cm]{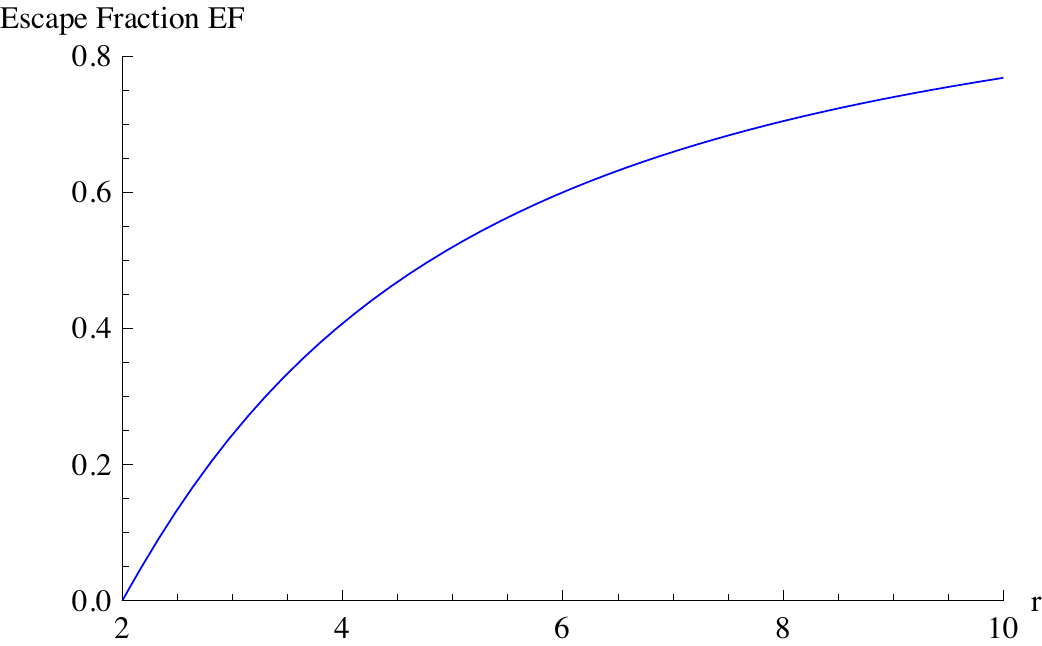}}
\vspace{0.5cm}
\caption{Escape fraction for a massless particle in the gravitational field of a Schwarzschild black hole as a function of $r$ for DM particles colliding with angular momentum $\al_1=2$ and $\al_2=-2$. \label{2min2}}
\end{figure}

In addition to the dependence of $EF$ on $r$ there is an implicit dependence on the initial angular momenta of the two colliding DM particles $\al_1$ and $\al_2$. We can plot this escape fraction for any combinations of $\al_1$ and $\al_2$ and in FIG.~\ref{2min2} we show $EF$ for $\al_1=2$ and $\al_2=-2$ as an example. With different combinations of angular momenta the size of the escape fraction does not change significantly.

As a consistency check, we can remove the boost from the calculation and consider massless particles produced at some position $r$ and allowed to have initial momenta in any direction in the LNRF. This is obviously a much simpler problem. Comparing with the calculation of the escape cone outlined in \cite{mtw} we find complete agreement.

We now move to the extremal Kerr case. The calculation for Kerr follows the same steps outlined for the Schwarzschild case. Let us repeat the reasoning for clarity: two DM particles collide at a given $r$ with two respective angular momenta. We have two frames: the LNRF frame, which we can think of conventionally as the lab frame, and the CMF of the collision. The two frames are related by a boost, which is defined by the radial position and the angular momenta of the colliding DM particles. Isotropic production of the massless particle in the CMF is again assumed. The conditions for escape, given by a range in $b$ and $\si$, as observed in the LNRF, are once again computed by considering the turning points of the massless particle. The boost is then applied to convert the momentum of the escaping particle to the CMF. This allows us to map the escape conditions in the LNRF, which are conditions on $b$ and $\si$, to a range of angles in the CMF. This is then divided by $2 \pi$ to obtain the escaping proportion of massless particles. Although the formula for the turning points is now more complicated, the reasoning is identical to the Schwarzschild case above. 

\begin{figure}[t]
\vspace{0cm}
\centerline{\includegraphics[width=8cm]{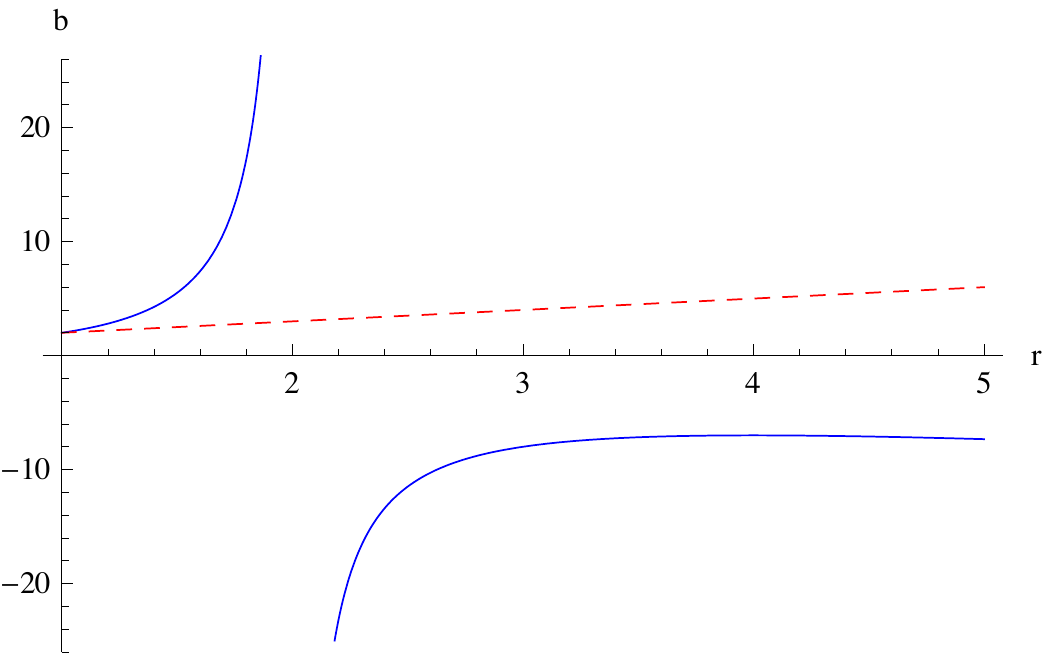}}
\vspace{0.5cm}
\caption{Plot of turning points for an extremal Kerr black hole in terms of $b$ as a function of $r$. The dashed (red) line is $b_+^{a=1}(r)$ whereas the solid (blue) curve is $b_-^{a=1}(r)$. \label{ae1turn}}
\end{figure}

The first step is to analyse the turning points of $\dot{r}$ once again. Using Eq.~(\ref{solb}) and setting $a=1$ we have the two solutions
 \bea
b^{a=1}_{+}(r)&=&r+1,\\
b^{a=1}_{-}(r)&=&-\frac{r^2-r+2}{r-2}.
 \eea
We have plotted these two solutions in FIG.~\ref{ae1turn}. Note that the conditions for escape are now more complicated, but the logic is the same.  If the particle has $\si=+1$ and it does not encounter a potential barrier, then it escapes. If it has $\si =-1$ and it encounters a potential barrier, then it bounces off and escapes. The solution, $b^{a=1}_{-}(r)$, has a maximum at $r=4$ with value $b^{a=1}_{-}(4)=-7$. The solution $b^{a=1}_{+}(r)$ has a lowest value of 2 which occurs at the horizon, $r=r_h$. Again, we will drop the superscript $a=1$ from what follows, as for the remainder of this section we will be concerned only with extremal Kerr. 

From this plot, we can read off the conditions needed for the massless particle to escape (note that here $b^{a=1}_{-}(r)$ refers to the lower (negative) branch of the curve, as displayed in FIG.~\ref{ae1turn}). They are for
\bea \label{con5}
r<4: \si&=&+1\;\;\;{\rm and}\;\;\;-7 <b <b_{+}(r);\\\label{con6}
 \si&=&-1\;\;\;{\rm and}\;\;\; 2 <b <b_{+}(r),
\eea
and for
\bea \label{con7}
r>4: \si&=&+1\;\;\;{\rm and}\;\;\; b_{-}(r) <b <b_{+}(r);\\\label{con8}
 \si&=&-1\;\;\;{\rm and}\;\;\;2 <b <b_{+}(r);\\\label{con9}
  \si&=&-1\;\;\; {\rm and}\;\;\;b_{-}(r) <b <-7\, .
\eea
Let us give the physical meaning of the escape conditions for $r<4$. In this case, a particle which is moving away from the horizon ($\si= +1$), will escape unless it hits the lower solid curve ($b_{-}(r)$) of FIG.~\ref{ae1turn}. To avoid that, it must have $b >-7$, which is the maximum of that curve. On the other hand, if it has a value of $b$ that is larger than the dashed line ($b_{+}(r)$), then it will hit that line and bounce back to the horizon: this explains the condition with $\si = +1$. For $r <4$ and $\si=-1$, the particle which is moving towards the horizon, then escape will occur if it hits the red line, thus having $b_{+}(r)>b > 2$. Similar reasoning yields the remaining conditions. 
We now need our final state massless particle momenta in the CMF. We follow the same procedure as before but now the initial DM momenta have the form

\bea\nonumber
Q_{i\hat{t}}(r, a=1, \al_i)&=&\frac{C-2 \alpha_i }{\sqrt{rC} (r-1)}\\\nonumber
Q_{i\hat{r}}(r, a=1,  \al_i)&=&-\frac{\sqrt{2 \left(1+r^2\right)-4 \alpha_i -(r-2) \alpha_i^2}}{\sqrt{r}(r-1)}\\
Q_{i\hat{\th}}(r, a=1, \al_i)&=& \alpha_i \sqrt{\frac{r}{C}},
\label{ae1comps}
\eea
where $C=r^3+r+2$.

The boost parameters are now set by the choices of $\al$ of the initial DM particles. Using the components in Eq.~(\ref{ae1comps}) and the boost, we can calculate the CMF momenta of the massless final state particles. These CMF momenta will be functions of $b$, $r$ and $\si$ and we can apply the conditions necessary for escape listed in Eqs.~(\ref{con5}, \ref{con6}, \ref{con7}, \ref{con8}, \ref{con9}).

Following again the procedure outline above we now want to construct dot products of the vectors
\beq\nonumber
P^{\prime}_{\phi}(r, b, \si)=\left[ P^{CM}_{\hat{r}}(r, b, \si, a=1),\; P^{CM}_{\hat{\phi}}(r, b, \si, a=1)\right].
\eeq
evaluated with parameters corresponding to the extreme escape values quoted in the escape conditions. The prime notation is there to remind the reader that we are evaluating this with $a=1$ (not to be confused with the tilde notation for the $a=0$ Schwarzschild case). Thus, we are able to obtain the angular range in the CMF where produced massless particles escape. 
This is in complete analogy to the Schwarzschild case above. 

The range of escape angles that corresponds to the range in the values of $b$ and $\si$ for $r<4$, as listed in Eqs.~(\ref{con5}, \ref{con6},) $\De \phi_{a=1}^{r<4}$, in the CMF reads
\begin{multline}
\De\phi_{a=1}^{r<4}(r)=
\\\abs{\arccos\;\left[\frac{P^{\prime}(r, \;b_{+}(r), \;+1).P^{\prime}(r, \;-7, \;+1)}{\abs{P^{\prime}(r, \;b_{+}(r), \;+1)}\abs{P^{\prime}(r, \;-7, \;+1)}}\right]}\\
+\abs{\arccos\;\left[\frac{P^{\prime}(r, \;b_{+}(r), \;-1).P^{\prime}(r, \;2, \;-1)}{\abs{P^{\prime}(r, \;b_{+}(r), \;-1)}\abs{P^{\prime}(r, \;2, \;-1)}}\right]}.
\end{multline}
For $r>4$, applying the conditions in Eqs.~(\ref{con7}, \ref{con8}, \ref{con9}) $\De \phi_{a=1}^{r>4}$ takes the form

\begin{figure}[t]
\hspace{-1cm}
\centerline{\includegraphics[width=8cm]{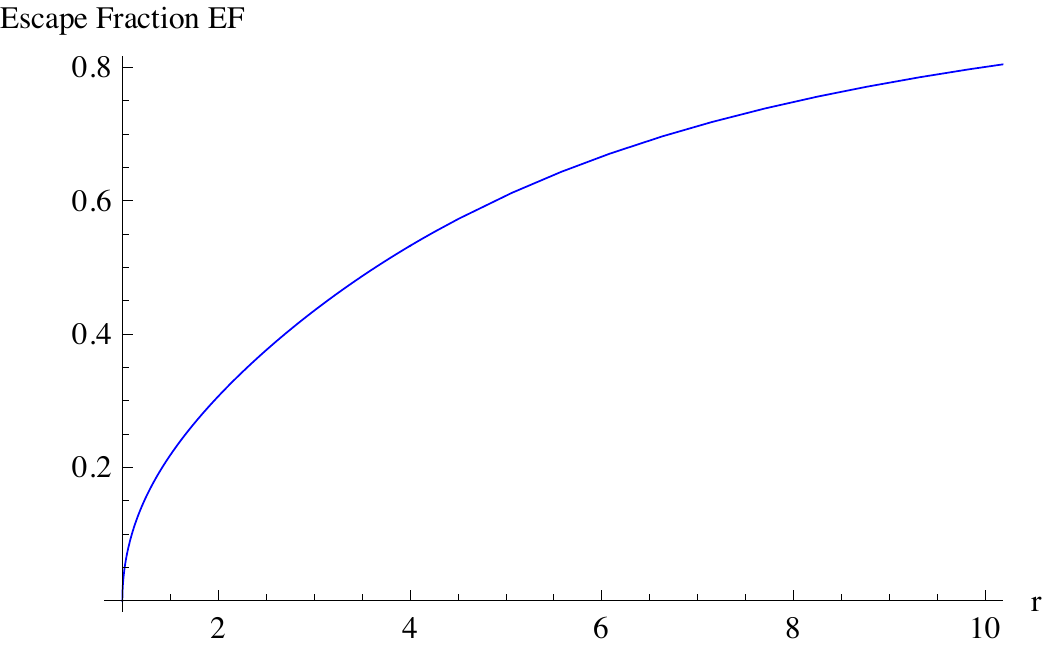}}
\vspace{0.5cm}
\caption{Escape fraction for a massless particle in the gravitational field of an extremal black hole as a function of $r$ for DM particles colliding with angular momentum $\al_1=2$ and $\al_2=-2 $. \label{escapekerr2-2}}
\end{figure}

\begin{multline}
\De \phi_{a=1}^{r>4}(r)=
\\\abs{\arccos\;\left[\frac{P^{\prime}(r, \;b_{+}(r), \;+1).P^{\prime}(r, \;b_{-}(r), \;+1)}{\abs{P^{\prime}(r, \;b_{+}(r), \;+1)}\abs{P^{\prime}(r, \;b_{-}(r), \;+1)}}\right]}\\
+\abs{\arccos\;\left[\frac{P^{\prime}(r, \;b_{+}(r), \;-1).P^{\prime}(r, \;2, \;-1)}{\abs{P^{\prime}(r, \;b_{+}(r), \;-1)}\abs{P^{\prime}(r, \;2, \;-1)}}\right]}
\\\abs{\arccos\;\left[\frac{P^{\prime}(r, \;-7, \;-1).P^{\prime}(r, \;b_{-}(r), \;-1)}{\abs{P^{\prime}(r, \;-7, \;-1)}\abs{P^{\prime}(r, \;b_{-}(r), \;-1)}}\right]}.
\end{multline}
The total $r$ dependent escape function is given by
\bea\nonumber
EF_{a=1}(r)&=&\frac{1}{2\pi}\left[\De\phi_{a=1}^{r<4}(r)\Theta(r-4)\right.\\
&&\left.+\De\phi_{a=1}^{r>4}(r)(1-\Theta(r-4))\right],
\eea
where $\Theta(r-4)=1$ for $r<4$ and zero otherwise. Again $EF_{a=1}(r)$ has an implicit dependence on the initial DM angular momentum $\al_a$ and $\al_b$. In FIG.~\ref{escapekerr2-2} we show a plot of the escape function as a function of $r$ for initial angular momentum $\al_a=2$ and $\al_b=-2$.

\begin{figure}[t!]
\centerline{\includegraphics[width=8cm]{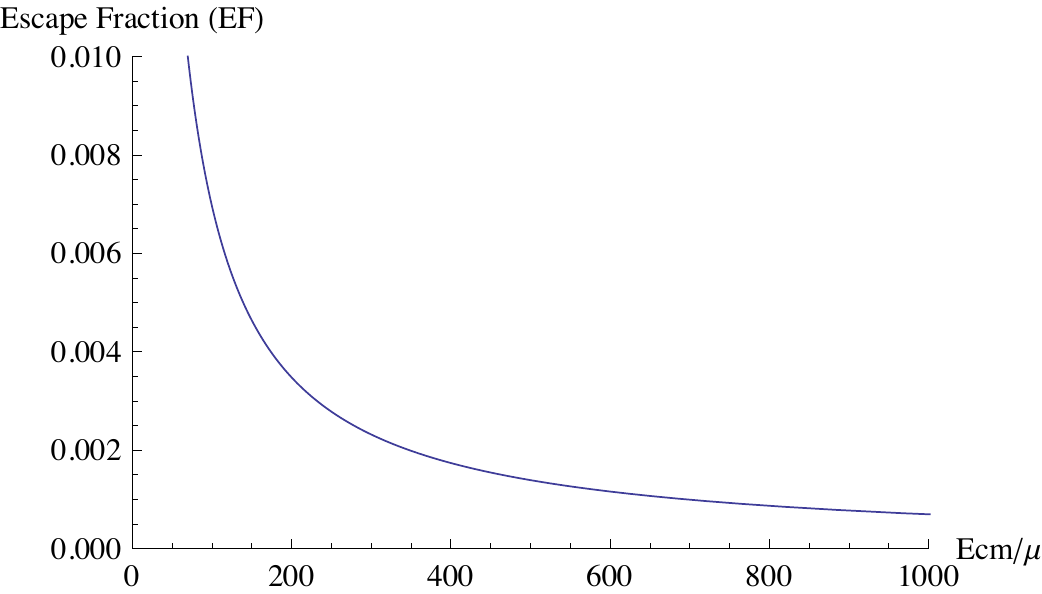}}
\vspace{0.5cm}
\caption{The escape fraction versus CMF energy for two DM particles colliding with angular momentum $\al_1=2$ and  $\al_1=-2$ in the gravitational field of a Kerr black hole with $a=1$. The energy is displayed in units of the DM mass $\mu$.  \label{evsesc}}
\end{figure}

Using this escape fraction, we can not only calculate the emergent flux of massless particles but we can plot the escape fraction vs the CMF energy of the collision. Doing so gives a clear measure of how likely it is that a massless particle that is produced in a very high energy collision will escape the black hole. We are able to do this because the CMF energy of the collision is purely a function of the radius and the angular momenta of the colliding DM particles, \cite{banados},
\begin{widetext}
\begin{multline}
\Big(E^{^{Kerr}}_{cm}\Big)^2=\frac{2\, m_0^2}{r (r^2 - 2r + a^2)} \Big(  2 a^2 (1 + r)-2 a (\alpha_2 + \alpha_1) - \alpha_2 \alpha_1 (-2 + r) + 2 (-1 + r) r^2   \\  -\sqrt{2 (a - \alpha_2)^2 - \alpha_2^2 r + 2 r^2} \sqrt{2 (a - \alpha_1)^2 - \alpha_1^2 r + 2 r^2}\, \Big).\label{EcmK}
\end{multline}
\end{widetext}
Given this relationship, we can construct the plot shown in FIG.~\ref{evsesc}, which displays the escape fraction versus CMF energy for two DM particles colliding with angular momentum $\al_1 = 2$ and  $\al_2 =-2$ and is evaluated in the extreme Kerr limit, $a=1$. It is noteworthy that even for CMF energies that are close to $1000$ times the DM mass, the escape function is almost $0.001$, which is non-negligible, especially if these collisions are taking place frequently which of course depends on the DM density in this region as well as the size of the cross section for DM collisions. To get an estimate for the size of the escape fraction for higher energies, a good approximation for the parametric behaviour plotted in FIG.~\ref{evsesc} is $EF\approx 0.7 E_{cm}^{-1} \mu$. This approximation is good for $E_{cm}>10\mu$, any lower and this approximate parametric dependence breaks down.

\section{The dark matter density profile}
\label{density}

Now that we have the escape fraction for a rotating black hole, we must turn to the DM density profile in the 
area immediately around the black hole, a region where reliable modelling of the DM dynamics is difficult. In what follows, we shall make some simplifying assumptions that enable us to make a reasonable estimate for that density. 
We consider an IMBH, taken to be of reference mass $10^5 M_5\rm M_\odot, $ as predicted to exist in the inner galaxy as failed building blocks for the central supermassive black hole, and possibly existing throughout the halo as failed satellites
\cite{BZS2005}. Here $M_\odot$ is the mass of the Sun and $M_5$ is a dimensionless number parameterising the absolute mass of the black hole. We caution that this is a hypothesis, and there may be other routes to building supermassive black holes. 

In addition, we assume that the DM consists of massive weakly interacting massive particles that froze out in the early universe. Doing so gives us rough ball park figure for the size of the annihilation cross section. Such particles are cold and form a density spike via adiabatic contraction during the formation phase of the IMBH by baryonic dissipation. For our purposes the details of the spike or any other part of the DM profile are irrelevant if they concern distances bigger than ${\cal{O}}(1)\times R_s$. All we need take from the profile is that the rising DM density as one approaches the horizon is tempered by its annihilation, so that in the immediate vicinity of the black hole an ``annihilation plateau'' is reached. The (larger) radius at which the plateau is reached 
is unimportant for our calculation because it is much larger than ${\cal{O}}(10)R_s$. On the other hand, the radius up to which the plateau continues (the inner radius of the plateau), we take to be the black hole horizon. Therefore, we assume that the DM density is essentially flat from the black hole horizon all the way up to values of the radius which are ${\cal{O}}(10) \times R_s$, where $R_s=G M_{BH}/c^2$. We emphasize that the details of the DM density profile at such radii are irrelevant for us, as we are interested in the most energetic particles resulting from annihilations taking place at radii that are 
 ${\cal{O}}(1) \times R_s$. The density in the annihilation plateau is given by $\rho_{pl} = m_{\chi}/\langle \sigma v \rangle t$, where $\langle \sigma v \rangle$ is the thermal-averaged annihilation rate per unit volume, $m_{\chi}$ 
 is the DM mass, and $t$ is the time-scale for spike formation. For details of the annihilation plateau, we refer the reader to \cite{BFTZ2009,BZS2005,BGZ1992, Gondolo:1999ef}. 
 
Although we now have an estimate for the {\emph{total}} DM density in the near-horizon region, we are still short of information as to the composition of the DM particles with regard to their angular momenta. 
This is of crucial importance because the energies of the emitted photons are set by the angular momenta of the colliding DM particles, and the very energetic collisions can occur for DM particles in plunge orbits. So what we must do is obtain an estimate of the proportion of particles in the plateau which are in fact within the plunge range of the black hole. The complexity of the problem means that this is beyond the scope of the paper, and, given that we are only interested in an order-of-magnitude estimate of the flux emitted in the immediate vicinity of the horizon, we simply assume a flat distribution in angular momentum. In addition, we assume that for the particles colliding near the horizon, the proportions of plunge and bound geodesics are comparable and so we take the absolute value of the density of both to be of order the height of the plateau. 

We remind the reader that the correct angular momentum decomposition of the DM particles in the plateau is likely to be very complicated, especially in the near-horizon region. Our aim in this analysis is to estimate the size of the total emergent flux originating from near horizon collisions.  It turns out that we are almost insensitive to the details of the angular momentum distribution as we are effectively integrating over the full range of angular momentum states. If we wanted to calculate the energy spectrum of the emerging massless particles the choice of the angular momentum distribution becomes more important with the potential to change the shape of the spectrum.

\section{Results}
\label{results}

\subsection*{The flux emitted in the near-horizon region}

In this section we put together the ingredients computed so far: the density in the near-horizon region and the escape fraction to calculate the flux. Going back to Eqs.~(\ref{annihilation}) and (\ref{flux}), for plunge orbits we can exchange the dependence on the velocities for a dependence on the angular momentum of the colliding particles. The phase space distributions, $n$, can in principle be recast  as functions of $x$ and $\al$. The $\al$ dependence encodes the distribution of angular momentum states which in general changes with $x$. This $x$ dependent distribution is unknown (as discussed above) and we make the approximation that we can factorize the $\al$ dependence from $n$ (such that this distribution is constant in $x$). We now assume that the remaining distribution function in $\al$ is flat across the allowed range in $\al$. In addition, the escape function $e(r,\alpha_1,\alpha_2)$ changes slowly with $\al_i$ and so we can treat it as a constant with respect to the integral over the angular momenta of the colliding particles. 

Further to this assumption we remind the reader that the escape fraction has been calculated for the equatorial plane only. To do the full volume integral we need to know how the escape function changes as we go off the equatorial plane. Instead of changing the form of the escape function to account for this variation (which is a task most easily completed numerically) we assume a spherical symmetry such that we can use the same escape fraction for the full volume.

We can convert the number density $n(\vec{x})$ to a mass density $\rho(\vec{x})$ by dividing by the rest mass $m_\chi$. For our purposes we set this density equal to the plateau density $\rho_{pl}$. The final form for the flux arriving at some distance $D$ away from the black hole simplifies to a spherically symmetric volume integral (a consequence of the assumed spherical symmetry of the density and escape function in the near-horizon region) and can be written as
\begin{eqnarray}\label{Flux}
\Phi&\approx&\frac{\sigma v R_s^3}{4 \pi m_\chi^2 D^2}\, \int_{r_h}^{r_1}\,  \rho_{pl}^2(r) \;e(r,\alpha_1,\alpha_2)\, dV, 
\end{eqnarray}
where $ \sigma v  $ is to be taken as some typical annihilation cross section, $ m_\chi $ is the mass of the DM particle, and $D$ is the distance to the black hole.  A further complication which we do not address here is that the cross section will in general be energy and therefore $r$ dependent.  As we wish to make a simple estimate of the flux we take the cross section to be constant in energy. A further simplification is that we assume that $ \sigma v $ is dominated by an s-wave term, which is velocity independent. In doing so $ \sigma v $ is assumed to be just a constant.
In general there will be velocity dependent terms and consequently we would replace $\si v$ with the equivalent thermally averaged cross section.

Note also that the form of the flux in Eq.~\ref{flux} is similar to the flux spectrum constructed in Ref.~\cite{BZS2005} with the addition of the escape function. Ref.\cite{BZS2005} deals with the flux spectrum $d\Phi/dE$ which requires an additional ingredient $ dN_\gamma/dE $, which is the number of secondary particles produced per annihilation. We are interested in the total flux of particles, rather than the flux spectrum (the latter is obviously very important, but its calculation is much harder in a relativistic setting).  We can trivially convert the flux spectrum to total flux by assuming that photons reaching us will have energies bunched around the mass of the DM particle, so that it is safe for our purposes to write $ dN_\gamma/dE =\delta(E-m_{\chi})$ and integrate trivially over energy to obtain the total flux in Eq.~(\ref{Flux}).

The escape fraction $e(r,\alpha_1,\alpha_2)$ was computed in Section~\ref{escapefraction}. It is relativistic, and so its convolution with a non-relativistic density is one of the major approximations in this work. As the escape function is slowly varying with $\al_i$  we may choose $\alpha_1,\alpha_2$ as we wish (if we knew the angular-momentum distribution of the plateau then we would have integrated over these two variables weighted by the distribution function in angular-momentum space). We choose one of $\alpha_1$ and $\alpha_2$ to be the critical angular-momenta corresponding to the highest-energy collisions. This choice will not alter the order of magnitude of the results below. The density $\rho_{pl}(r)$ is simply the {\emph{mass}} density of DM particles that are on plunge orbits. As outlined above, we will use $\rho_{pl}(r)= \frac{m_{\chi}}{\sigma v t}$, again implicitly assuming that the proportion of plunge particles is of the same size as non-plunge particles. 

Note that we are integrating from the horizon $r_h$ up to an upper radius $r_1$ which is a few horizon lengths. The factor of $R_s^3$ is necessary for converting the integral from $R_s=1$ units to SI units. Plugging everything in we are left with 
\begin{eqnarray} \label{eqn:flux}
\Phi&=&\frac{R_s^3}{2 \langle \sigma v \rangle  D^2 t^2}\, {\cal{I}}(r_1,a) \, , \nonumber \\
{\cal{I}}(r_1,a) &=& \int_{r_h}^{r_1}\, r^2 \,e(r,\alpha_1,\alpha_2)  \,  dr \, ,\nonumber \\
&&
\end{eqnarray}
where we have isolated the dimensionless integral ${\cal{I}}(r_1,a)$ inside which is found {\emph{all}} of the dependence on the rotation of the black hole. This dependence on the rotation parameter $a$, which enters into the limits of the integral as well as the integrand, is obviously rather complicated. Computing this integral for extremal Kerr and for Schwarzschild, with $r_1=r_h+4$ for both (and with $\alpha_1=2,\alpha_2=-2$ for Kerr and  $\alpha_1=4,\alpha_2=-4$ for Schwarzschild), we obtain ${\cal{I}}_K\approx 20$ and ${\cal{I}}_S \approx 40$. For comparison, we note that if we put 1 in place of the escape fraction we obtain the values ${\cal{I}}_K\approx 40$ and ${\cal{I}}_S \approx 70$. 

We can now give an estimate of the expected emergent flux. We use the following fiducial values for our various parameters: we set the mass of the black hole using $M_5$ and set it equal to $M_0 = 40$ in units of $10^5$ solar masses. We use $ (\sigma v)_0= 10^{-28}  \rm cm^2 s^{-1}$, set the distance between us and the black hole to $D_0=10 \;\rm kpc$ and assume that the growth time-scale of the black hole is $t_0=10^{10}$ years. The total flux is then given by:
\begin{eqnarray}
\Phi & = & \Phi_0 
\left( \frac{ \sigma v }{(\sigma v)_0} \right)^{-1} 
\left( \frac{D}{D_0} \right)^{-2}\left( \frac{t}{t_0} \right)^{-2}\left( \frac{M_5}{M_0} \right)^{3} {\cal{I}}(r_1, a)\, ,  \nonumber \\
&&
\end{eqnarray}
where  
$ \Phi_0=3.41 \;\mbox{km}^{-2} \mbox{year}^{-1} $. Note that by writing the flux in this fashion we have placed all of the dependence on the geometry into the integral ${\cal{I}}(r_1, a)$. We see that IMBHs can be bright DM annihilation sources, and this has been explored and constrained in the literature \cite{aharonian2008}. For the case of extremal Kerr, integrating up to $r_1=5$, we obtain the flux to be 
$70 \;\mbox{km}^{-2} \mbox{year}^{-1}$. 
This is well within the reach of planned neutrino detectors. If we integrate only up to $r_1=1.1$ for extremal Kerr, thereby including the emergent flux of massless particles coming from the most energetic annihilations, we obtain an expected number of  $0.26 \;\mbox{km}^{-2} \mbox{year}^{-1}.$ Futuristic  large area high energy neutrino/particle detectors in space such as OWL ($http://owl.gsfc.nasa.gov/$)
would have an effective detection area solid angle efficiency product  in excess of $10^5\rm km^2sr$ and could detect such signals over a few days (admittedly only above $10^{19}\rm eV$). 
We note below that Penrose boosting may allow such energies to be achievable at infinity.   

While the actual flux from an IMBH depends on its distance from us and on the other reference parameters, which are  uncertain by several orders of magnitude, we note  that our calculation of a finite escape fraction which is  weakly dependent on the black hole angular momentum means that even detection of a single event might open a window on new physics, attainable only via particle collisions  at extreme CMF energies. Our final result should be viewed as an order of magnitude estimate
of the flux received on earth, which we  find to be the same with or without black hole rotation. On the other hand, given that very high (even arbitrarily high) energies
can be achieved in the annihilations around a rotating black hole, while the Schwarzschild geometry only allows {\emph{maximal}} CMF energies of $2\sqrt{5} m_{dm}$,
our results  imply that the spectrum received from a rotating black hole should typically contain signatures of highly energetic products. A next step is  to compute the spectrum
as a function of $a$ to confirm this expectation. We leave this for future work.

\section{open issues}

We obtained above an order-of-magnitude estimate for the blue-shifted flux emitted by DM annihilations in the near-horizon region of an extremal Kerr black hole. We made several approximations and assumptions along the way that made the calculation tractable. We assumed that the escape function calculated purely on the equatorial plane was sufficient, restricting the colliding DM particles and the escaping particles to motion only on the equatorial plane. We also assumed that the DM ``spike" around the black hole was non-relativistic and spherically symmetric, and that it was flat right down to the black hole horizon. In our view, the former of these two concerns is unlikely to be a source of large errors, because the escape fraction goes from zero to one regardless of the dimensionality or the geometry. It is very easy to check using our results above that the change from Schwarzschild to Kerr makes very little difference to the escape function. Moreover, some work we have done, and upcoming work \cite{williams}, shows that going to three-dimensions does not change the results significantly, thus corroborating our intuition. The concern about the DM density profile is rather more serious, because the density enters the flux quadratically. Several effects need to be studied more rigorously, chief among them being the effect of the rotating geometry on the DM spike, as well as the feasibility of a more accurate modelling of the spike close to the black hole horizon in a fully relativistic fashion. The advantage of such a construction is that it should naturally yield the composition of angular momentum in the DM spike, which is crucial for the flux and its spectrum.  

We should mention here that a reliable computation of the flux spectrum requires another ingredient: an explicit particle physics model of the DM and its interactions. This is necessary because it determines the energy dependence of the differential cross section of the annihilation diagrams, which, coupled with the geometry-dependent CMF energy of the collisions, feeds directly into the differential flux spectrum. The flux spectrum is therefore expected to be a far more revealing signature than the total flux, as it should contain clues as to the nature and composition of the DM spike. For example, we naively expect the spectrum to bunch around the DM mass, which in itself would be an exciting observation.

There are further necessary refinements of our work that we would like to comment on. The computation of the CMF energy in the Kerr field reference \cite{banados} neglected back-reaction effects, which were studied in \cite{jacobson, berti}. This issue deserves further attention.  Finally, supermassive black holes, the nearest one of which is at our Galactic Centre, should not be ruled out as possible sources of annihilating  DM. Even after partial destruction by  a complex merging history, something which is not necessarily an issue very close to $R_s$, cusps, albeit with softer profiles , are regenerated by dynamical processes
\cite{merritt2007}.  Finally, one of the most intriguing aspects, that we reserve for future work, is the exploration of the Penrose process \cite{penrose1969}, which may allow the escaping annihilation products to tap the rotational energy of the Kerr black hole, provided that the collisions occur within the black hole ergosphere. This effect was briefly considered in \citep{grib}, but the potential remains to be fully explored.\\

\section{Acknowledgments}

The work of MB was partially supported by the JS Guggenheim Memorial Foundation and  Fondecyt (Chile) Grant \#1100282 and \# 1090753.
BH thanks Christ Church College, Oxford, for financial support, and CERN, where some of this work was carried out. SMW thanks the Oxford physics department for hospitality and the Higher Education Funding Council for England and the Science and Technology Facilities Council for financial support under the SEPNet Initiative.


\end{document}